Homology Modeling and Analysis of Human Proteins related to Disease

# Structural Investigations into Shwachman Bodian Diamond Syndrome SBDS using a Bioinformatics Approach

[1]Babu A. Manjasetty, [2]Sunil Kumar, [3]Andrew P. Turnbull, [4]Niraj Kanti Tripathy
[1]Proteomics & Bioinformatics Research Group, Research & Industry Incubation Center,
Dayananda Sagar Institutions, Bangalore 560 078 Karnataka, INDIA
babu.manjasetty@gmail.com
[2]Institute of Life Sciences, An Autonomous Institute of Department of Biotechnology,
Govt. of India, Nalco Square, Bhubaneswar 751023, Orissa, INDIA
[3]Cancer Research Technology Ltd., Birkbeck College, University of London, London, WC1E 7HX, United Kingdom
[4]Department of Zoology, Berhampur University, Berhampur 760 007, Orissa, INDIA

## Abstract

**The functional correlation of missense mutations which cause disease remains a challenge to understanding the basis of genetic diseases. This is particularly true for proteins related to diseases for which there are no available three dimensional structures. One such disease is Shwachman Diamond syndrome SDS OMIM 260400, a multi system disease arising from loss of functional mutations. The *Homo sapiens* Shwachman Bodian Diamond Syndrome gene *h*SBDS is responsible for SDS. *h*SBDS is expressed in all tissues and encodes a protein of 250 amino acids *SwissProt accession code Q9Y3A5*. Sequence analysis of disease associated alleles has identified more than 20 different mutations in affected individuals. While a number of these mutations have been described as leading to the loss of protein function due to truncation, translation or surface epitope association, the structural basis for these mutations has yet to be determined due to the lack of a three-dimensional structure for SBDS. Moreover, SBDS is a highly conserved gene of unknown function, however, biochemical studies suggest that SBDS may be involved in RNA metabolism or ribosome assembly. Here, we describe a homology model for *h*SBDS which is based on the crystal structure of AF0491 from *Archaeoglobus fulgidus* PDB code 1P9Q that shares 24 percent sequence identity and 49 percent similarity with *h*SBDS. This model provides insights into how the different mutations are involved in translation, truncation, protein protein interactions, protein stability and function.**

## 1. 1INTRODUCTION

Structural proteomics (SP) is a global and rapidly growing research domain following the recent successes in genome sequencing and technical breakthroughs in macromolecular crystallography (MX) and nuclear magnetic resonance (NMR) [1]. MX is a very powerful technique allowing the structures of proteins, nucleic acids and their complexes to be determined at atomic resolution. Worldwide SP initiatives have developed and implemented technologies that facilitate automated, high-throughput three-dimensional protein structure determination [2]. The main focus of these initiatives is to determine the structures of a diverse set of proteins which span as many protein families as possible. With the advent of advances in protein structure analysis, experimentally determined three-dimensional structures of representative members of large protein families are being generated at an ever increasing pace. For example, the protein structure initiative (PSI) produced over 1,300 structures during its pilot phase (year 2000-2005) and, to date, approximately 3,000 structures have already been deposited in the protein data bank (PDB)[3] during the second phase (year 2006-2010). This success is also evident in the increased contribution from other world-wide SG/SP centers in terms of the number of structures being deposited into the PDB [4].

Increased structural coverage of protein families with high biomedical importance enables the exploration of structure-function relationships and aides computational modeling. Homology modeling is a powerful technique that greatly enhances the value of experimentally determined structures by using the structural information from one protein to predict the structures of related proteins. Protein families in which at least one experimental structure has been determined can act as a template for modeling other family members (provided the level of sequence identity is at least 25%). Current methodologies and technologies for homology modeling can produce structural models with highly accurate backbones and an intermediate level of accuracy for side chains and loops [5,6,7]. Given that the number of experimental structures available is far lower than the huge number of sequenced proteins, high quality models can be produced for a proportion of these proteins of unknown structure and reliable models can be generated for up to 60% of the remaining proteins encoded by a particular genome. Homology modeling plays an important role in filling the huge gap between experimentally determined structures and sequenced proteins, especially for the structures of proteins related to human diseases which prove difficult to determine experimentally. Thus, the vast majority of biologists are increasingly reliant on homology models of proteins of interest. Additionally, the protein structures of disease-related proteins, experimental or predicted, are of great research interest to both experimental and computational biologists. Their high value in medicine and human health stems from the fact that they provide informative clues as to how they function in the cell and a molecular snapshot of the disease process which aides rational drug design [8]. This is particularly true for diseases caused by alterations of proteins for which no three-dimensional structure is available. With the help of homology modeling, disease-related mutations can also be studied at the atomic level. To date, thousands of proteins have been implicated in various human diseases. Frequently, mutations in these proteins have been identified in patients suffering from a particular disease. Mutational data often allows these diseases to be studied at the molecular level by generating the protein structure either through experimental procedures or by homology modeling.

In this report, a homology model for the *Homo sapiens* Shwachman-Bodian-Diamond Syndrome (*h*SBDS) gene is presented. SBDS is responsible for Shwachman-Diamond syndrome (SDS; OMIM 260400), a multi-system disease arising from loss-of-function mutations [9]. SDS is a rare autosomal recessive marrow failure syndrome associated with exocrine pancreatic insufficiency and leukemia predisposition. Bone marrow failure typically manifests with neutropenia, but anemia, thrombocytopenia, or aplastic anemia may also develop. Other clinical features include skeletal, immunologic, hepatic and cardiac disorders [9-12].

The *SBDS* gene is ubiquitously expressed and encodes a protein of 250 residues (*SwissProt accession code Q9Y3A5*). The function of this protein is unknown and it has no primary sequence similarity to any other protein or structural domain that could indicate a possible function. However, biochemical studies suggest that SBDS protein may be involved in RNA metabolism or ribosome assembly. Many of the disease-associated mutations and truncations have also been identified. The model presented here may serve as a valuable reference to improve the understanding of the molecular basis of this disease and to provide a detailed view of the disease-related mutations within the structural domains of the protein.

## 2. MATERIALS AND METHODS

*Template Search and Sequence Alignment*: The protein sequence of Shwachman-Bodian-Diamond syndrome (Accession no. AAN77490) of *Homo sapiens* was retrieved from the NCBI database. A Basic Local Alignment Search Tool (BLAST) [13] search was performed against the Protein Data Bank (PDB) [14] using default parameters to find suitable templates for homology modeling. Based on the maximum identity with high score and low e-value, 1P9Q at 2.0 Å resolution was selected as the most appropriate template for homology modeling. The sequence identity and similarity are 24% and 49% between the template (1P9Q) and the target sequence respectively. The *Clustal-W* (http://www.ebi.ac.uk/clustalw) [15] program was used to produce the sequence alignment between *h*SBDS and 1P9Q.

*3D Structure Generation:* The academic version of MODELER (http//:www.salilab.org/modeler) [16] was used for three-dimensional structure generation based on the information obtained from the sequence alignment. Out of 20 models generated by MODELLER, the one with the best G-score in PROCHECK [17] and the best VERIFY3D [18] profile was subjected to energy minimization. Using the parameters as a distance-dependent dielectric constant $\varepsilon = 1.0$ and non-binding cutoff of 14 Å, CHARMM [19] force field and CHARMM-all-atom charges, initially a 600 step steepest descent algorithm was used to remove close Van der Waals contacts, followed by an 800 iteration conjugate gradient minimization until the energy showed stability in sequential repetition. All hydrogen atoms were included during the calculation. The energy minimization described above was started with the core main chain and then extended to all the core side chains. All calculations were performed using the ACCELRYS DS Modeling 2.0 (Accelrys Inc. San Diego, CA 92121, USA) software suite. During these steps, the quality of the initial model improved significantly. STRIDE [20], which uses hydrogen bond energy and main chain dihedral angles to recognize helix, coils and strands, was subsequently used to predict the secondary structure composition of the modeled SBDS. The weighted root mean square deviation (RMSD) of the modeled protein was calculated using the combinatorial extension (CE) algorithm [21]. The modeled structure was then superimposed on the crystal template without altering the coordinate systems of atomic position in the template. The RMSD for the superimposition is 2.1Å. The residue profiles of the three-dimensional models were further checked using VERIFY3D [18].

PROCHECK [17] analysis was performed in order to assess the stereo-chemical properties of the three-dimensional models and Ramachandran plots were calculated. No residues reside in the disallowed regions of the Ramachandran plot.

## 3. RESULTS AND DISCUSSION

### Homology Model

The X-ray structure of an SBDS homologue in *Archaeoglobus fulgidus*, AF0491 [22], was used as the structural template (PDB 1P9Q) for the homology modeling protocol. The sequence alignment between human SBDS (*h*SBDS) and *Archaeoglobus fulgidus* AF0491 (*Af*SBDS) is shown in Figure 1. All conserved residues are coloured grey. The alignment is characterized by some insertions and deletions in loop regions. Since the first 9 N-terminal residues and the last 9 C-terminal residues (residues 241–250) in SBDS do not have corresponding residues in 1P9Q, the modeling was carried out from residues 10-240, followed by a rigorous refinement of the model by means of Energy Minimization (EM - see the Materials and Methods section for the protocol used). The three-dimensional conformation of the predicted model of SBDS is presented in Figure 2. The stereo-chemical quality of the modeled SBDS was evaluated using PROCHECK. The main chain phi/psi angles for 94.9% of residues fall in the most favored regions, 4.7% of residues lie in the additionally allowed regions and 0.5% fall in the generously allowed regions. No residues are present in unfavourable regions of the Ramachandran plot (Figure 3). The overall PROCHECK G-factor for the homology modeled structure was – 0.05. Thus, statistical analysis suggests that the backbone conformation of the predicted model of SBDS is almost as good as that of the template. VERIFY3D provided the plot which indicates that the final protein model is correct. The secondary structure elements of SBDS were predicted using the program STRIDE, which uses hydrogen bond energy and main chain dihedral angles to recognize helix, coils and strands which are well defined in the modeled SBDS structure.

### Overall Structure

The core features of the three dimensional SBDS homolog (*Af*SBDS) structure are retained in the *h*SBDS model (Figure 2)[22, 23]. The subunit structure of *h*SBDS is organized into three domains: an N-terminal, central and C-terminal domain. The N-terminal domain was first identified in the *Af*SBDS structure and displayed no structural similarity to other protein folds in the PDB. Therefore, this domain represented a novel fold and was termed the Fungal, Yhr087wp, Shwachman domain (FYSH). The central domain comprises $\alpha$-helices arranged in helix-turn-helix motifs suitable for RNA or DNA binding [24]. The C-terminal domain resembles a ferrodoxin-like fold which commonly exists in a large number of proteins [25]. The structural superposition of the $C_\alpha$ atom positions of the *h*SBDS homology model and *Sf*SBDS (Figure 4) gives an RMSD of 2.1Å with a significant Z-score of 6.8. Structural differences stretching throughout the model represent conformational flexibility between the homologues.

### Implications for SDS

The wide occurrence of SBDS homologues provides opportunities to glean insights into the molecular basis of this disease [26]. The structure of *Af*SBDS was recently determined and indicated that the tertiary structure of this protein is arranged in three domains, comprising a novel N-terminal FYSH domain followed by two domains that are highly conserved in other structural domains [22,23,26]. On the other hand, biophysical and biochemical functional data from model organisms clearly provides clues to functional conservation in family members and a phylogeny study reveals that the characteristics of a functionally coherent group of SBDS proteins is highly suggestive of a role in translation and/or RNA metabolism [27]. The structural and functional analyses of SBDS orthologues, coupled to its localization within the nucleus, suggest an involvement in RNA processing [22,28].

More than 20 mutations have been identified in *h*SBDS which are highlighted in reverse type in Figure 1. The functional consequences of SBDS mutations have not yet been characterized, however, mutations in the affected patients result from gene conversion between SBDS and an SBDSP pseudogene [29]. Most mutations are present in the N-terminal FYSH of *h*SBDS suggesting that this domain plays an important physiological role. The most important point mutation, C31W, is located within the hydrophobic core of the FYSH domain (Figure 5). Strikingly, C31 has the potential to form a disulfide bond with another mutated residue, C84, to enhance structural stability. The structural consequences of the dual C31 and C84 mutations will be to destabilize the protein which will adversely affect protein function. Therefore, these mutations are responsible for protein stability and will eventually cause SDS.

A detailed understanding of the functions of *h*SBDS is not known, however, based on the structural features identified in the modeled structure, one can speculate potential roles. In this regard, residues located on the surface of SBDS proteins are likely to be important for interacting with partner proteins and any mutation of these residues will affect the surface electrostatic potential. Some of the missense mutations such as E44G, K118N, E28A and R100A, which are implicated in SDS disease, are localized on the surface of *h*SBDS and have been mapped onto the modeled structure (Figure 6). Based on proteomics data, an SBDS-interactome has been developed and SBDS binding partners with diverse molecular functions, notably components of the large ribosomal subunit and proteins involved in DNA metabolism, have been reported [30]. The in-depth comparison of experimental data with the structural data will provide further insights into SDS.

## 4. CONCLUSION

To date, thousands of genes (proteins) have been identified to be associated with various human diseases. Frequently, mutations in these proteins have been identified in patients suffering from a particular disease. Homology modeling is a valuable tool to help to understand how these mutations cause disease. Here, a homology model of *h*SBDS has been generated using a bioinformatics approach. Our analysis provides insights into the distribution of the mutations implicated in SDS and highlights how they can influence protein stability and alter the surface electrostatic potential. We plan to extend the current study to the prediction of the structures of complexes of disease proteins. The resulting complex models would provide an important resource for studying the mechanism of disease.

## BIOGRAPHY

**Dr. Babu A Manjasetty** is Professor & Chief Scientist at Research & Industry Incubation Center (RIIC), Dayananda Sagar Institutions (DSI), Bangalore. He leads the Proteomics & Bioinformatics Research Group at RIIC. Prior to joining DSI, he worked at Case Proteomics Center, Brookhaven National Laboratory, New York, USA & as an Alexander von Humboldt (AvH) Fellow, Germany at Max-Delbrück Center for Molecular Medicine, Berlin, Germany. He obtained his Ph.D in Physics from the University of Mysore under University Grant Commission (UGC) fellowship.

**Sunil Kumar** is Information Officer at Institute of Life Sciences, An autonomous Institute of Department of Biotechnology, Govt. of India, Bhubaneswar. He is looking after the Bioinformatics Centre. He holds masters degree in Biotechnology and Advanced Diploma in Bioinformatics. He is pursuing Ph.D in Bioinformatics from Berhampur University, Berhampur, India. His research interest is Structural Bioinformatics and Sequence analysis.

**Dr Andrew Turnbull** is Principal Scientist of structural biology at Cancer Research Technology Ltd., London, United Kingdom. He set up and leads the new crystallography unit responsible for determining the structures of pharmaceutically important protein-ligand complexes to aid CRT's small molecule drug discovery programs. Dr Turnbull obtained his PhD in Biochemistry from the University of Sheffield with work on amino acid dehydrogenases.

**Prof. Niraj Kanti Tripathy** is Professor in Zoology at Berhampur University, Berhampur, Orissa (India) He is having 35 years of research and Teaching experience. He worked in Europe for 5 years.

```
 hSBDS MSIFTPTNQI RLTNVAVVRM KRAGKRFEIA CYKNKVVGWR SGVEKDLDEV 50
SfSBDS --------MV SLDKAVIARL RKGGEEFEVL VDPYLARDLK EGKEVNFEDL 42

 hSBDS LQTHSVFVNV SKGQVAKKED LISAFGTDDQ TEICKQILTK GEVQVSDKER 100
SfSBDS LAAEEVFKDA KKGERASVDE LRKIFGTDDV FEIARKIILE GEVQITAEQR 92

 hSBDS HTQLEQMFRD IATIVADKCV NPETKRPYTV ILIERAMKDI HYSVKTNKST 150
SfSBDS REMLEAKRKQ IINFISRNTI DPRTNAPHPP SRIERALEEA KVHIDIFKSV 142

 hSBDS KQQALEVIKQ LKEKMKIERA HMRLRFILPV NEGKKLKEKL KPLIKVIESE 200
SfSBDS EAQVKDIVKA LKPILPLKFE EMEIAIKIPP EHTGRAISAL YNFGGVTR-E 191

 hSBDS DYGQQLEIVC LIDP--GCFR EIDELIKKET KGKGSLEVLN LKDVEEGDEK FE 250
SfSBDS EWQRDGSWIC VMRIPSGMYG DLMDLLGKVA KGEALTKVLR RIG------- -- 234
```

Figure 1. Sequence Alignment between Human SBDS (*h*SBDS) and *Af*SBDS. Identical residues are highlighted in grey (total of 24% identity). The positions of the human SBDS-associated mutations are highlighted in reverse type.

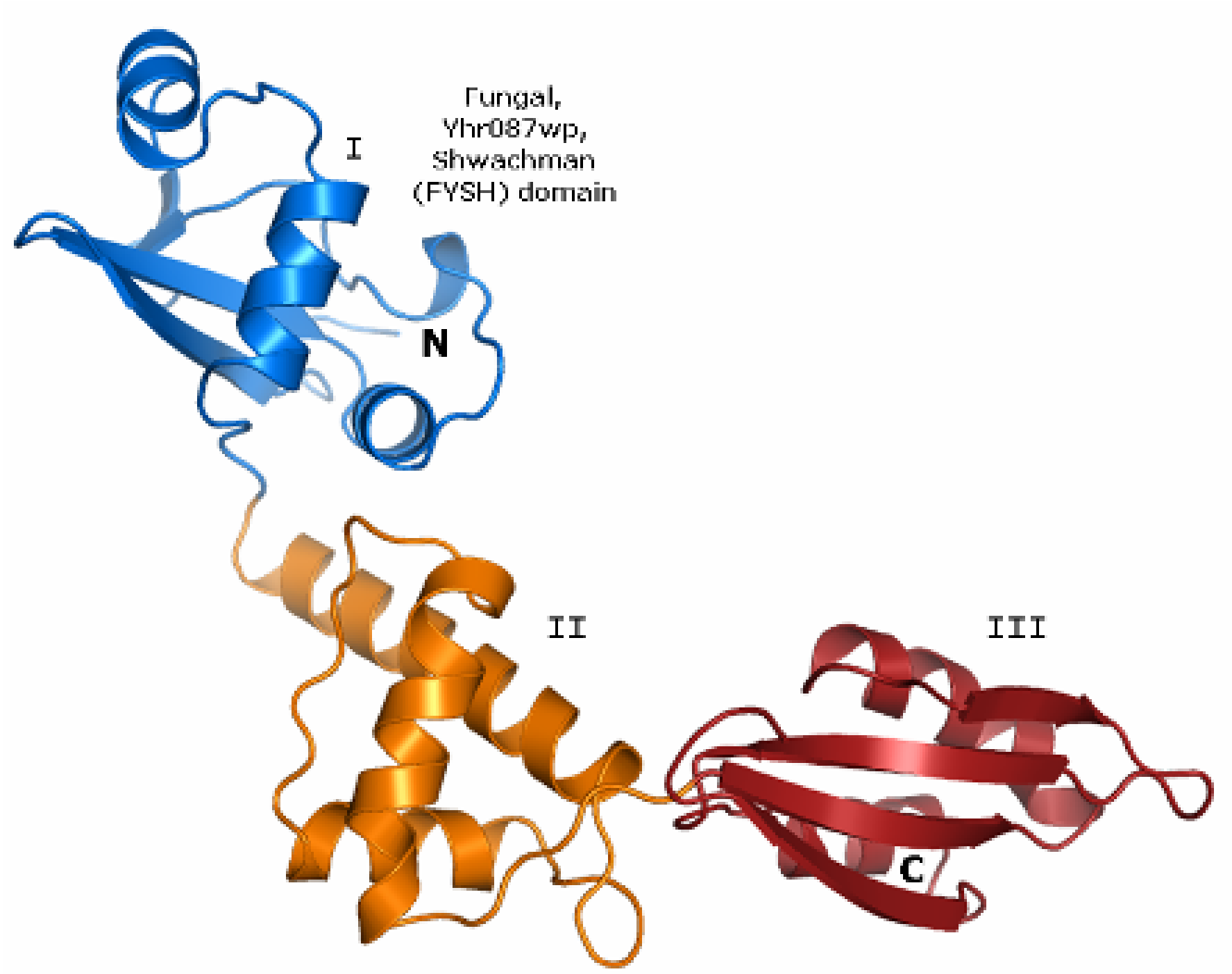

Figure 2. Schematic representation of the modeled SBDS structure. Figure prepared using the program PyMOL.

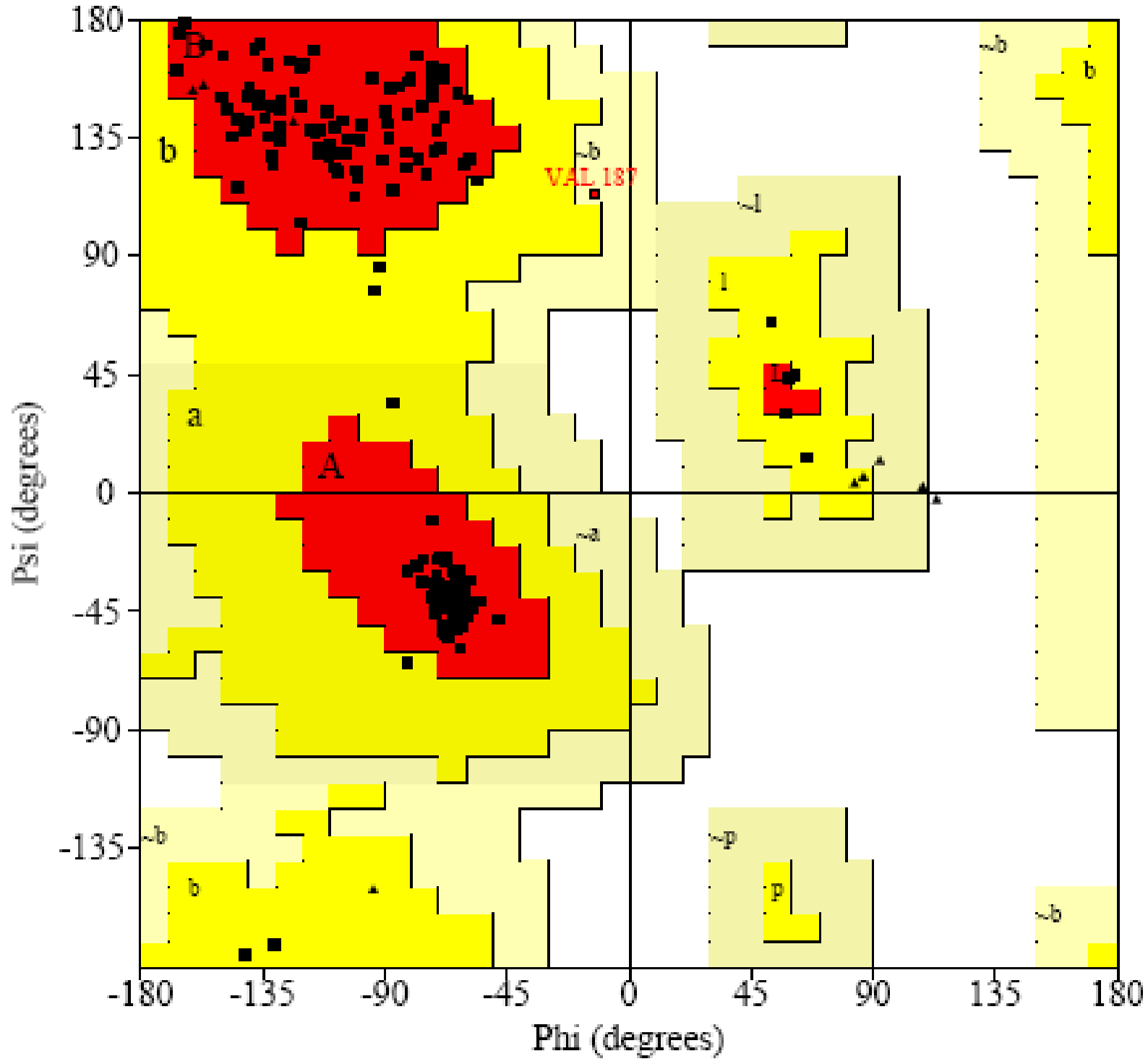

Figure 3. Ramachandran plot for the *h*SBDS model**.**

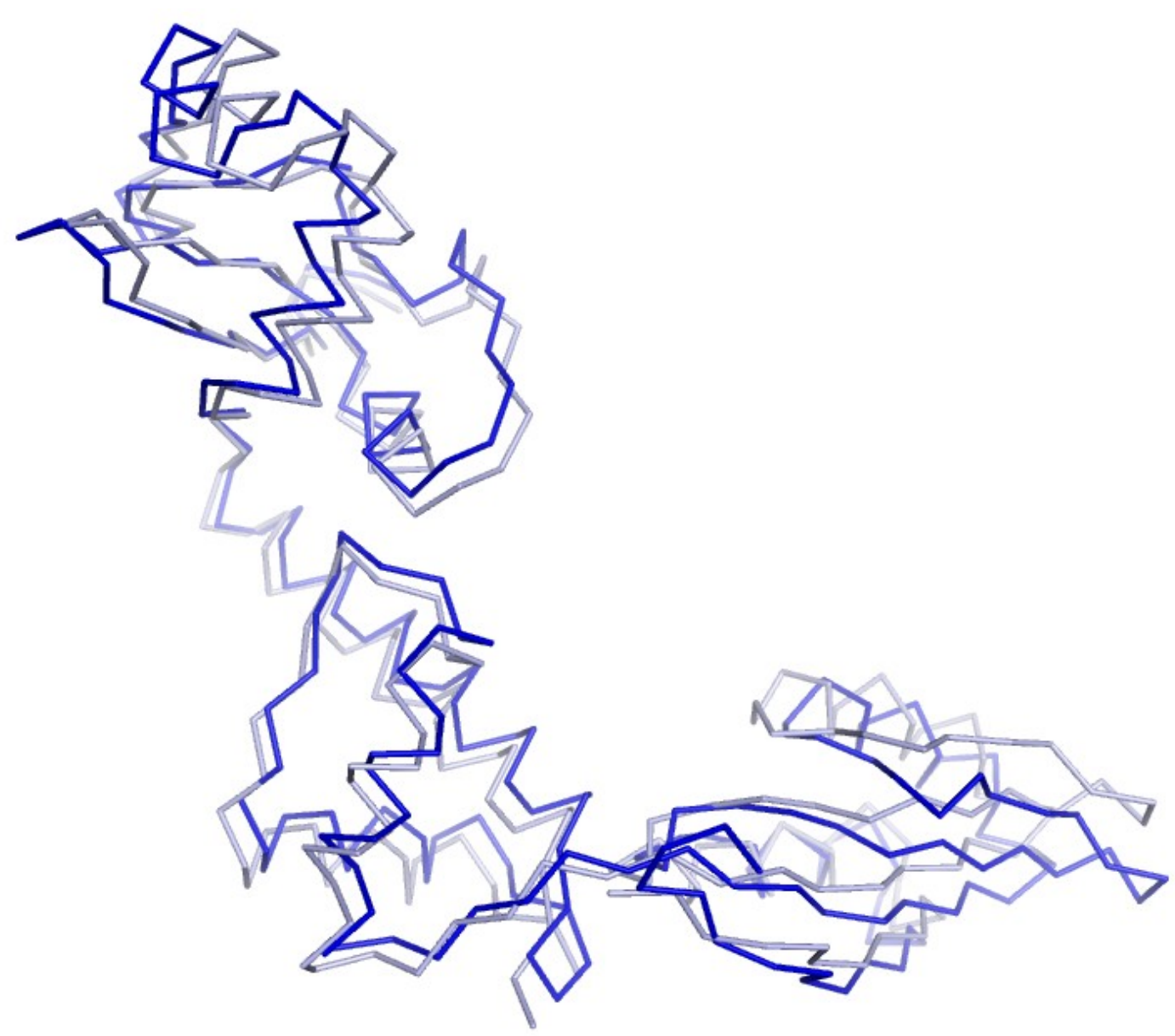

Figure 4. Cα trace of the *h*SBDS model (dark blue) superposed on *Af*SBDS (light blue). Figure prepared using the program PyMOL.

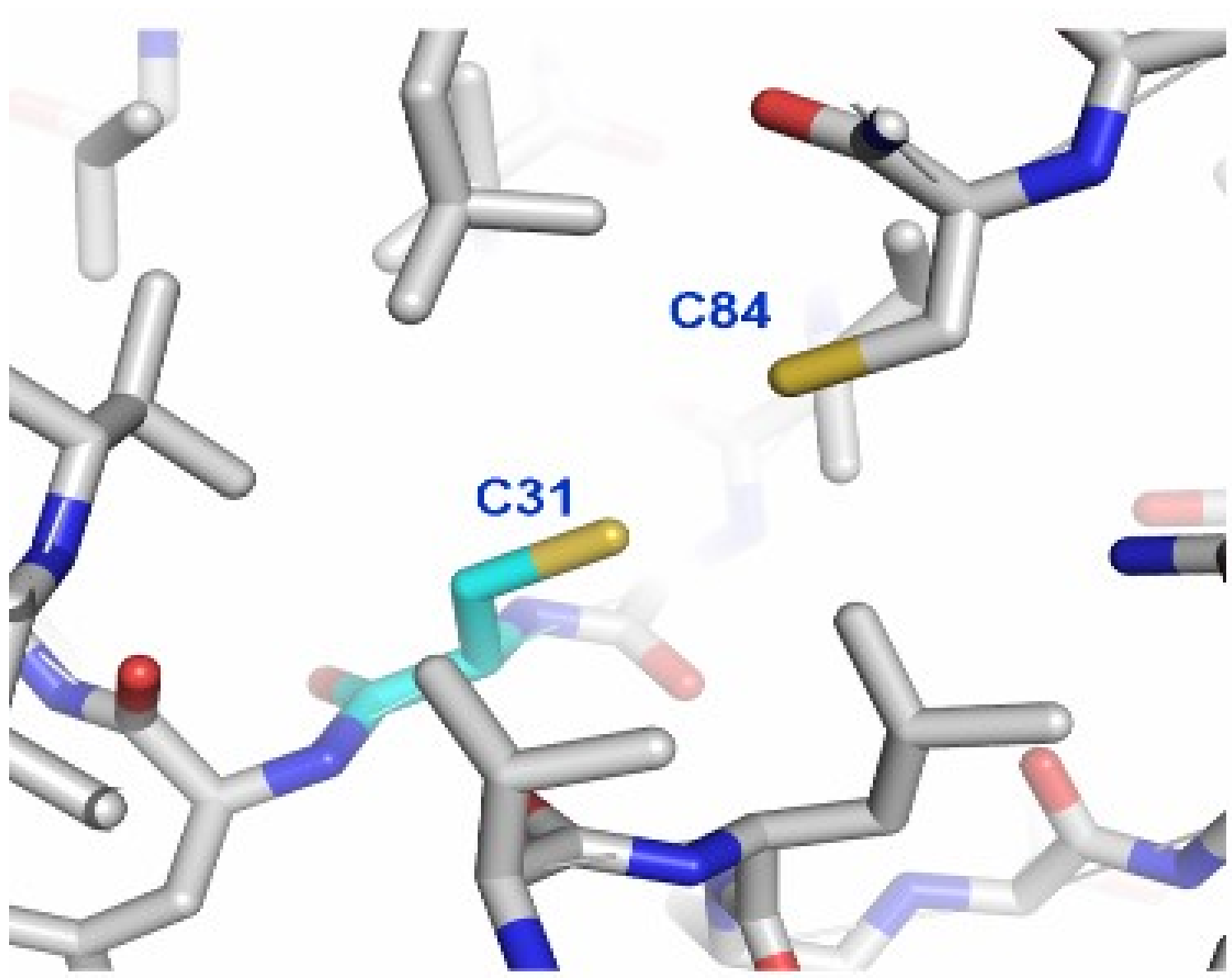

Figure 5. The key point mutation, C31W, is located within the hydrophobic core of the FYSH domain and resides within close proximity to C84

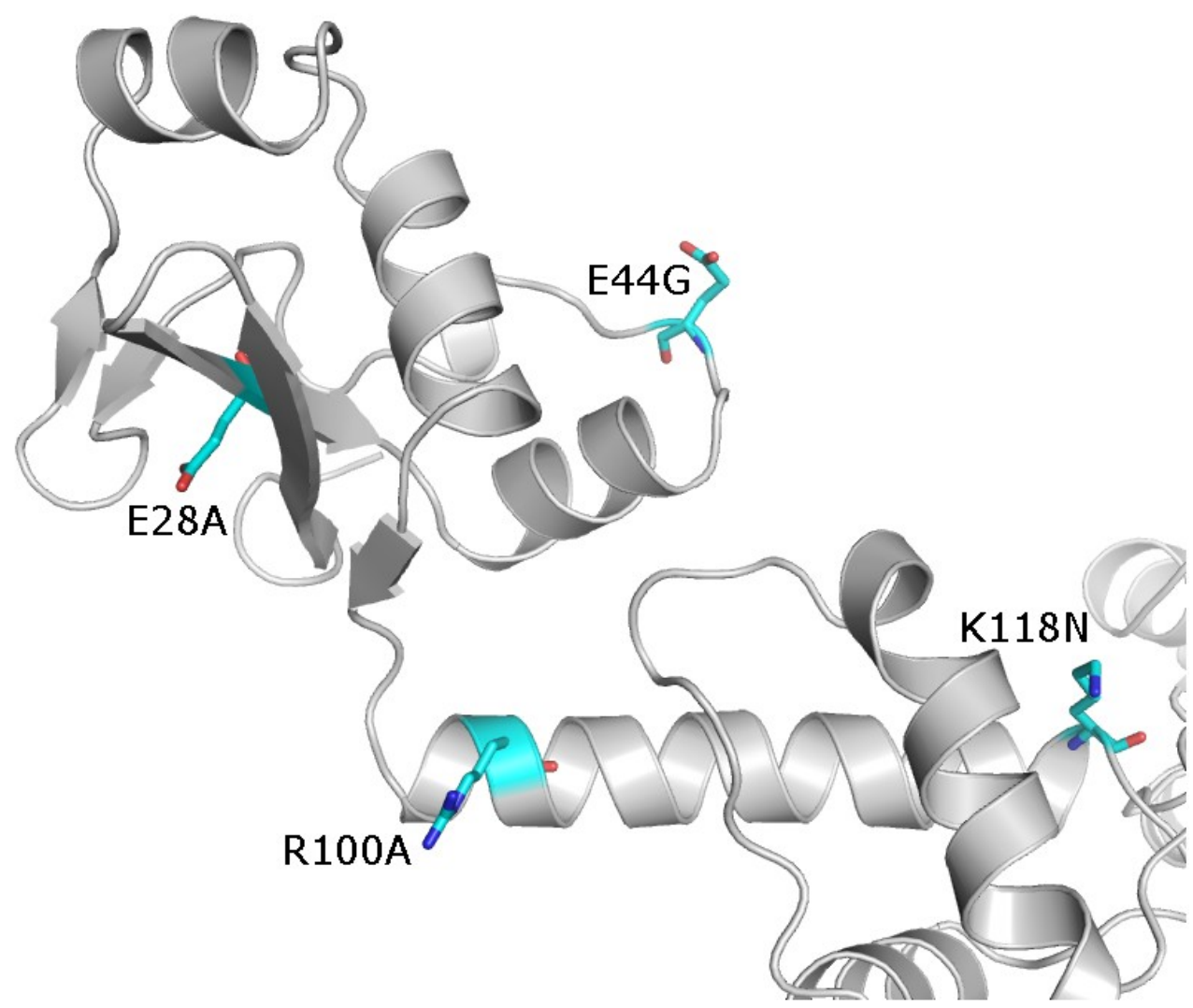

Figure 6. Some of the surface localized mutations involved in interactions highlighted in cyan